\documentclass[10pt, a4paper]{article}
\usepackage{jheppub}
\pdfoutput=1
\usepackage{graphicx,microtype,appendix}
\usepackage{booktabs,multirow}
\usepackage[dvipsnames]{xcolor}
\usepackage[colorlinks=true]{hyperref}
\usepackage[capitalise]{cleveref}

\allowdisplaybreaks

\title{Displaced Searches for Light Vector Bosons at Belle II}
\author[a]{Triparno Bandyopadhyay\;,}
\author[b,c,d]{Sabyasachi Chakraborty\;,}
\author[c]{Sokratis Trifinopoulos\;.}

\affiliation[a]{Department of Theoretical Physics, Tata Institute of Fundamental Research, Mumbai 400005, India\;,}
\affiliation[b]{SISSA International School for Advanced Studies, Via Bonomea 265, 34136, Trieste, Italy\;,}
\affiliation[c]{INFN - Sezione di Trieste, Via Bonomea 265, 34136, Trieste, Italy\;,}
\affiliation[d]{Department of Physics, Florida State University, Tallahassee, FL 32306, USA\;.}

 \emailAdd{triparno@theory.tifr.res.in}
 \emailAdd{sachakr@sissa.it}
  \emailAdd{sokratis.trifinopoulos@ts.infn.it}
 
\preprint{{\raggedleft SISSA 04/2022/FISI,\\ TIFR/TH/22-3. \par}}

\abstract{
With a design luminosity of 50~ab\textsuperscript{-1} and detectors with tracking capabilities extending beyond 1 m, the Belle II experiment is the perfect laboratory for the search of particles that couple weakly to the Standard Model and have a characteristic decay length of a few centimetres and more. We show that for models of dark photons and other light vector bosons, Belle II will be successful in probing regions of parameter space which are as of now unexplored by any experiment.  In addition, for models where the vector boson couples sub-dominantly to the electron and quarks as compared to muons, e.g. in the $L_\mu-L_\tau$ model, Belle II will probe regions of mass and couplings compatible with the anomalous magnetic moment of muon. We discuss these results and derive the projected sensitivity of Belle II for a handful of other models. Finally, even with the currently accumulated data, $\sim 200$ fb$^{-1}$, Belle II should be able to cover regions of parameter space pertaining to the X(17) boson postulated to solve the ATOMKI anomaly.
}

\begin{document}
\maketitle

\section{Introduction 
    \label{sec:intro}
} 
The physics of weakly coupled light particles has gained a lot of traction in recent times. This is motivated in equal parts by the interest in weakly interacting dark sectors, the persistence of different anomalies at low energies, and the establishment of dedicated experimental programs designed to look for signatures of new light degrees of freedom. One such example is the Belle II~\cite{Belle-II:2018jsg} experiment that has recently started to collect data from the collision of \(e^+ e^-\) beams provided by the SuperKEKB collider~\cite{Akai:2018mbz}. The Belle II experiment, like its  predecessors, the BaBar~\cite{BaBar:1998yfb} and the Belle~\cite{Belle:2012iwr} experiments, is expected to push the high intensity frontier of particle physics to new limits. Light (\(\sim\) 1 GeV) new physics can produce a plethora of striking signatures at the Belle II experiment, of which, in this paper, we concentrate on tracks that can be reconstructed to coincide at vertices away from the interaction point (IP), i.e., displaced vertices. There has been a lot of attention in recent times on displaced vertex searches at the intensity frontier. For example, strong projected limits are obtained in the context of Belle II for sterile neutrinos~\cite{Dib:2019tuj}, heavy QCD axions~\cite{Bertholet:2021hjl} based on the two loop amplitude computed in Ref.~\cite{Chakraborty:2021wda}, and dark photons (DP)~\cite{Duerr:2019dmv,Ferber:2022ewf}. For lighter DP masses, strong constraints have been obtained~\cite{Galon:2022xcl} using displaced DP searches at the MUonE experiment~\cite{Abbiendi:2022oks}, recently. 

From a theoretical perspective, light vector bosons are motivated by many new physics~(NP) scenarios. Arguably, the most popular of these models are those where the light vector boson acts as a portal between the Standard Model~(SM) particles and some new dark sector~\cite{Boehm:2003hm,Fayet:2004bw}. The DP model is the most well studied example of such portals to dark sectors. In these models, only dark sector particles have couplings to the new vector  boson, the DP (\(A^\prime\)). The \(A^\prime\) field strength has kinetic mixing with the SM photon \cite{Okun:1982xi,Galison:1983pa, Holdom:1985ag} which induces interactions between the \(A^\prime\) and the SM fermions. Another class of models of interest are those where the anomaly free global symmetries of the SM are gauged, viz., \(L_\mu-L_\tau\), \(L_e-L_\mu\), \(L_e-L_\tau\), and \(B-L\) \cite{Fayet:1980rr,Foot:1990mn, He:1990pn, He:1991qd}. Here, \(L_i\) is the lepton number corresponding to the \(i\) type lepton, \(B\) and \(L\) are the flavor-blind baryon number and lepton number respectively. Also, \(L_i-L_j\)/\(B-L\) can act as portals to dark sectors that carry lepton number and/or baryon number \cite{Kadastik:2009cu, Frigerio:2009wf,Heeck:2015qra, Bandyopadhyay:2017uwc, Foldenauer:2018zrz}. 
    
Light vector bosons, especially the \(L_\mu-L_i\) type vectors, have recently gained a lot of attention in the context of the muon and the electron \(g-2\) anomalies \cite{Ma:2001md, Pospelov:2008zw,Heeck:2011wj, Buras:2021btx}. The \((g-2)_\mu\) anomaly has persisted across different experiments for about a couple of decades~\cite{Muong-2:2006rrc, Aoyama:2020ynm}, with a new measurement quoting a tension of 4.2\(\sigma\) between the SM and the experimental numbers~\cite{Muong-2:2021ojo}. Although, a recent lattice result questions the validity of this discrepancy~\cite{Borsanyi:2020mff}, the anomaly holds as of now and the experimental precision is expected to be improved at upcoming facilities at the Fermilab \cite{Muong-2:2015xgu} and J-PARC \cite{Abe:2019thb}. Although the muon anomaly is more striking, there also exists an anomaly for the electron magnetic moment. The status of this anomaly is intriguing as two different computations with the fine-structure constant (\(\alpha_\mathrm{EM}\))  measured from two different sources, the Caesium (Cs) \cite{Parker:2018vye,Aoyama:2017uqe} and Rubidium (Rb) \cite{Davoudiasl:2018fbb, Morel:2020dww} atoms, give results on either side of the SM value. 

Given the theoretical interest in these particles, there have been numerous experimental searches for them. The non-observance of such particles has translated to constraints in the mass--coupling parameter space. To put our work in context, we briefly discuss the relevant experiments which have already given constraints using visible and invisible channels\footnote{See Ref.~\cite{Graham:2021ggy} for a comprehensive review.}. To probe visible channels, one requires $M_{A^{\prime}}>1$ MeV (\(>2m_e\)). The visible channels can be further classified into prompt and displaced, i.e., cases where the vector boson decays immediately after production and cases where it propagates for a finite distance before decaying. For a particle to decay away from the interaction point, it needs to have comparatively small decay widths. Hence, displaced vertex searches typically bound regions of small couplings and masses compared to prompt searches. We now categorize the existing experimental searches for an $A^{\prime}$-like boson at different experimental setups: 

\begin{enumerate}
    \item {\bf Searches in $e^+ e^-$ Colliders}: 
        In lepton colliders, the most prominent production mode for the \(A^\prime\) is the $t$-channel annihilation of electron positron pairs, $e^+ e^-\to A^\prime \gamma$~\cite{Fayet:2007ua}, followed by the decay of the $A^\prime$ to a pair of leptons/hadrons. The background is mostly dominated by the irreducible $s$-channel process $e^+ e^-\to\ell^+\ell^-\gamma$. So far, the most stringent bounds on the parameter space from \(e^+e^-\) collisions come from the BaBar experiment~\cite{BaBar:2014zli,BaBar:2016sci, BaBar:2017tiz}. The projected limits from Belle~II~\cite{Belle-II:2018jsg, Jho:2019cxq} are supposed to improve this bound a few times. For both these experiments, the center of mass (CoM) energy is around $10$ GeV. On the other hand, the KLOE experiment~\cite{Anastasi:2015qla, KLOE-2:2018kqf}, operating at the $\phi$ meson resonance ($\sim 1$ GeV) with $2.5$ fb$^{-1}$ data, probes significant areas in the low mass region. There are bounds on the \(A^\prime\) parameter space from the data collected by the BESIII detector, at the BEPCII collider, as well \cite{BESIII:2017fwv}. We do not show these bounds in our results as the BaBar bounds for the same masses are stronger. 
    \item {\bf Searches at Hadron Colliders:}
    In hadron colliders, the main production mode for light DPs with $M_{A^\prime}\lesssim 0.5$ GeV is from meson decays, e.g., $\pi^0\to A^\prime\gamma$ $\left(\eta\to A^\prime\gamma\right)$ etc. For $0.5~\text{GeV}<M_A^\prime<1~\text{GeV}$, mixings with the $\rho$, $\omega$, and $\phi$ mesons become important. Whereas, the Drell-Yan process with $q\bar q\to A^\prime$ gradually takes over for heavier masses. Till date, the strongest bounds from hadron colliders come from the LHCb collaboration, from both prompt and displaced searches~\cite{LHCb:2017trq, LHCb:2019vmc}. The dominant background for the prompt channel comes from irreducible SM Drell-Yan processes. On the other hand, the largest contribution for displaced searches stem from photon conversions. Rejecting dilepton vertices from the primary vertex can significantly reduce the background to a negligible amount~\cite{LHCb:2019vmc}.
    \item{\bf Searches at Meson factories:} 
    Dedicated experiments like NA48/2~\cite{NA482:2015wmo}, NA62~\cite{NA62:2019meo} etc. looked for $\pi^0\to A^\prime \gamma$ followed by the prompt decay of the DP to lepton pairs. The source of neutral pions are from Kaon decays. These experiments provide the strongest bounds in the light $M_{A^\prime}$ region. We only show the bounds from the NA48/2 experiment, as it is stronger than those from NA62.
    \item{\bf Searches in Proton Beam Dumps:} 
    Proton beam dump experiments like CHARM~\cite{CHARM:1985anb,Tsai:2019buq}, NOMAD~\cite{NOMAD:1998pxi, Gninenko:2011uv} etc.\ had a proton beam colliding against a fixed target (Cu and Be respectively). The hit at the target by the proton beam generates a plethora of mesons which subsequently decay to \(\gamma A^\prime\), $\pi^0\to A^\prime\gamma$, $\eta^\prime\to A^\prime\gamma$ etc.\ The strongest bound, however, comes from the \(\nu\)-calorimeter I (NuCal) experiment that looks at data from proton beam collisions on Iron at the IHEP-JNR neutrino detector \cite{Blumlein:1990ay,Blumlein:1991xh, Tsai:2019buq}.  
   
    \item{\bf Searches in Electron Beam Dumps:} 
    We look at the following electron beam dump experiments: E137~\cite{Bjorken:1988as}, E141~\cite{Riordan:1987aw},KEK~\cite{Konaka:1986cb}, Orsay~\cite{Davier:1989wz}, APEX~\cite{APEX:2011dww}, and A1 at MAMI~\cite{Merkel:2014avp}. These experiments exploit the bremsstrahlung process $e Z \to e Z A^\prime$, where \(Z\) is the SM neutral vector boson. The DPs are primarily produced in the forward region with subsequent decays to boosted leptons in the same direction. These experiments use a shielding to absorb SM particles located after their beam dump targets. The dimensions of the shielding restricts the $A^\prime$ flight of distance in the lab frame, which in turn sets a limit on the mass-coupling parameter space. For the models where the \(A^\prime\) does not couple to hadrons, the E137 bound is the strongest among the beam dump bounds. The E141 bound and the Orsay bound also covers complementary regions of the parameter space. Similarly, limits from NA64~\cite{NA64:2019auh} probe the region in-between the NA48 and the NuCal/E137 bounds. 
\end{enumerate}
We emphasise, the NA64 fixed target bound comes from a search of the hypothetical X(17) vector boson. This particle has been proposed as a solution to the \(\sim\) 17 MeV ATOMKI anomaly \cite{Feng:2016ysn}. Multiple experiments at the Tandetron accelerator of ATOMKI have reported anomalies in the angular correlation spectra of \(e^+e^-\) pairs in the decays of both excited\(~^8\)Be and\(~^4\)He \cite{Krasznahorkay:2018snd, Krasznahorkay:2021joi}.  As a result of this, different experiments across the globe have dedicated analyses to look for this particle, including NA64. We show later that Belle II displace searches should be sensitive enough to be competitive with the NA64 bound at current luminosities.

In our case, the exclusion limits that we draw entirely depend on displaced signatures of \(A^\prime\)-like vector bosons. To be specific, we look at displaced vertex signatures accompanied by a monophoton. For the decay of the \(A^\prime\), we look at dilepton (\(e+\mu\)) final states only. We checked for hadronic final states but did not find any sensitivity of the detector to displaced vertices for the mass ranges where the hadronic (or \(\tau\)) final states become kinematically accessible. We draw projected limits for two different luminosities. These are the Belle II design luminosity of 50~ab\textsuperscript{-1} and a luminosity approximately equal to the amount of data the collaboration is expected to have collected, \(\sim\) 200~fb\textsuperscript{-1}. We find that at the design luminosity, the experiment should be able to bound a large portion of the DP parameter space in a region that is as of now not excluded by any experiment. The region, that Belle II is expected to constrain, falls between the limits from prompt searches at BaBar and NA48/2 on the one hand and those from displaced searches at NuCal, E137 etc.\ on the other. We also find that even at current luminosities, the limits would be competitive with experiments like NA64 and E141, which aim to bound similar parameter regions as Belle II.

The paper is organised as follows: In \cref{sec:eff}, we discuss the tracking capabilities of the Belle~II detector and also discuss in detail our estimates for the displaced vertex reconstruction efficiencies at the trackers. In \cref{sec:cs}, we discuss the relevant theoretical calculations and go on to use the DP model to establish our methodology in obtaining the projected limits. In \cref{sec:model_zprime}, we extend our analysis to some other models of \(Z^\prime\) vector bosons. The models we discuss are the \(L_i-L_j\), \(B-L\) and the protophobic model~\cite{Fayet:1986rh}. Finally, in \cref{sec:conclusion}, we conclude.

\section{Track reconstruction at Belle II}
\label{sec:eff}
The Belle II experiment collects data from the events generated by the collisions of the SuperKEKB \(e^+e^-\) beams~\cite{Akai:2018mbz}, with asymmetric energies of \(E_+\simeq 4\) GeV and \(E_-\simeq 7\) GeV respectively, giving a CoM energy of \(\sqrt{s} \simeq 10.6\) GeV. The Belle II detector complex consists of several sub-systems placed in a cylindrical structure around the beam line. For this work, we are entirely interested in the detectors with tracking capabilities. These are, in order of increasing radial distance from the beam line, a silicon-pixel detector (PXD), a silicon-strip detector (SVD) (together called the vertex detector (VXD)), and a central drift chamber (CDC)~\cite{Belle-II:2018jsg}.

Our goal in this paper is to limit the parameter space of light \(A^\prime\)-like bosons, that couple very weakly to the SM. Specifically, we are interested in the regions of parameter space for which such a boson will travel for some distance before decaying to its daughter particles within the detector. A particle produced at the IP and coming out at an angle \(\vartheta\), wrt the electron beam, that decays after travelling a distance \(l\), will travel \(r= l\sin\vartheta\) in the radial direction, measured perpendicular to the beam-line, and \(z=l\cos\vartheta\) measured along the direction of the beam-line. If this decay happens within the fiducial volume of interest then the event is accepted with some efficiency, and rejected otherwise. The fiducial volume is defined not only by the physical dimensions of the tracking detectors but also by the requirements of efficient track reconstruction of the decay products of the $A^\prime$ inside said volume. As for the dimensions of the relevant components of the detector, the VXD spans a radial distance of 14 mm -- 14 cm, while the CDC spans from the end of the VXD to about 1.13 m. Both the VXD and the CDC have a \(\theta\)
coverage of:  
\begin{equation}
    \label{eq:fid_vol_eta}
    17^\circ < \theta < 150^\circ\,,
\end{equation}
which we incorporate as angular cuts in our analysis.  Also, following Ref. \cite{Dib:2019tuj}, we impose the following cuts on \(r\), \(z\):
\begin{equation}
    \label{eq:fid_vol}
    10\; \mathrm{cm} < r < 80\; \mathrm{cm}\,;\quad 
    -40\; \mathrm{cm} < z < 120\; \mathrm{cm}.
\end{equation}
We take \(r_\mathrm{min}\) = 10~cm to reject backgrounds arising from prompt tracks, e.g., decay products of \(K^0_S\) and \(\Lambda\), and tracks originating from secondary interactions of particles with the detector material. The upper limit, \(r_\mathrm{max}\) = 80~cm ensures that the decay happens deep enough inside the CDC for the resulting tracks to be reconstructed, with the CDC extending till 113 cm. In this paper, we work in the CoM frame, therefore, we scale all axial lengths by the appropriate boost factor, \(\gamma=1.04\), between the lab frame and the CoM frame. Belle II has a dedicated algorithm for the reconstruction of vertices away from the collision point, the \(V_0\)-like particle reconstruction algorithm. The efficiency of this algorithm, as demonstrated in Ref.~\cite{Belle-II:2018jsg}, falls with falling \(p_T\) of the $A^\prime$. To incorporate this in our analysis, we impose a conservative \(p_T\) cut:
\begin{align}
    \label{eq:pt_cut}
    p_T>0.9\; \mathrm{GeV},
\end{align} 
which can be relaxed with increase in the efficiency of \(V_0\) reconstruction.

For a decay vertex that falls within the fiducial volume, there is an efficiency of reconstruction of the tracks of the daughter particles. This reconstruction is performed using track-fitting and track-finding algorithms \cite{Belle-II:2018jsg}. A complete experimental analysis that employs these tools is beyond the scope of this work. Instead, to estimate the falling efficiency of vertex reconstruction with increasing radial distance from the beam-line, we use the linearly falling function of radial distance (\(r\)) that has been used in the analysis of
Ref.~\cite{Bertholet:2021hjl}: 
\begin{align}
    \label{eq:eff_r}
    \mathcal{E}(r) &=
        \frac{r_\mathrm{max}-r}{r_\mathrm{max}-r_\mathrm{min}},
\end{align}
where \(r_\mathrm{min/max}\) have been defined in \cref{eq:fid_vol}.

The probability of an \(A^\prime\), after being produced at the IP of the electron and positron beams, travelling a distance \(l\) from the IP before decaying is given by the exponentially falling decay distribution:
\begin{align}
    \label{eq:len_dist}
    P(l)&= \frac{e^{-l/\lambda}}{\lambda}.
\end{align}
The characteristic length \(\lambda\) defines this decay distribution and is determined from the mean-life, \(\tau\) of the particle and its boost:
\begin{align}
    \label{eq:l_half}
    \lambda&= \frac{|\vec{p}\;(M_{A^\prime})|}{M_{A^\prime}}\;c\tau(M_{A^\prime}, g^\prime),
\end{align}
where, \(c\) is the velocity of light in vacuum and \(|\vec{p}|\) is the magnitude of the three-momentum of the particle. Therefore, for a particular point in the \((M_{A^\prime},\,g^\prime)\) parameter space, the probability of identification of a displaced vertex, corresponding to an event with an \(A^\prime\) coming out in the direction \(\vartheta\) is a convolution of the efficiency given in \cref{eq:eff_r} with the decay length distribution in \cref{eq:len_dist}. Then, with \(r = l\sin\vartheta\) and \(z = l\cos\vartheta\), the vertex reconstruction efficiency (VRE) distribution, as a function of \(\vartheta\), is: 
\begin{align}
    \begin{split}
    \mathcal{E}(\vartheta)&= \frac{1}{\mathcal{N}}
        \frac{1}{r_\mathrm{max}-r_\mathrm{min}} 
        \int_{l_\mathrm{min}}^{l_\mathrm{max}} (r_\mathrm{max}-l\,
        s_\vartheta) \;
        e^{-l/\lambda}\; \overline{\Theta}(l) \; dl\,,
    \\\mathrm{where}&\quad 
        \overline{\Theta}(l)=\Theta\left(l\,s_\vartheta-r_\mathrm{min}\right)\,
        \Theta(r_\mathrm{max}-l\,s_ \vartheta)\,
        \Theta(l\,c_ \vartheta-z_\mathrm{min})\,
        \Theta(z_\mathrm{max}-l\, c_\vartheta)\;,
    \\\mathrm{and}&\quad l_\mathrm{min}= r_\mathrm{min};\;
    l_\mathrm{max}=\sqrt{r_\mathrm{max}^2 + z_\mathrm{max}^2},\quad
    \mathcal{N}= \int_{l_\mathrm{min}}^{l_\mathrm{max}}\;  
        e^{-l/\lambda}\; \overline{\Theta}(l) \; dl.
    \label{eq:fin_eff}
    \end{split}
\end{align}
In the above equation, we have used the shorthand notations \(s_\vartheta\equiv \sin\vartheta\) and \(c_\vartheta\equiv\cos\vartheta\). The efficiency under consideration is only for reconstruction of displaced tracks. The final signal estimate also includes an overall particle identification efficiency, as mentioned later.

\begin{figure}[htpb]
    \includegraphics[scale=0.98]{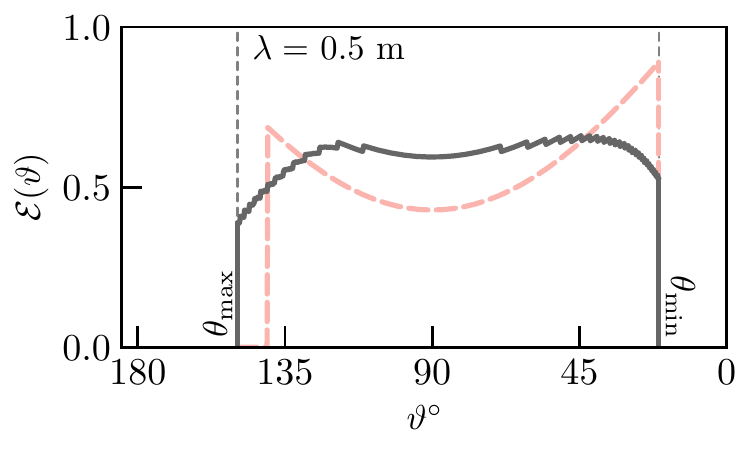}
    \includegraphics[scale=0.98]{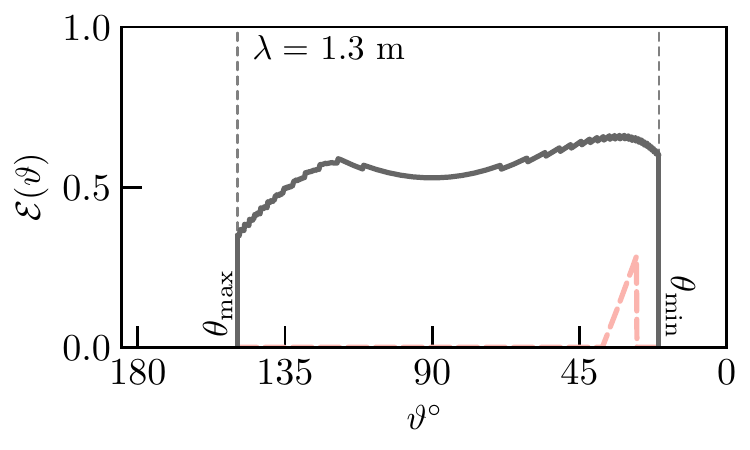} 
    \caption{
        We plot, using solid black lines, the vertex reconstruction efficiency distributions as a function of \(\vartheta\) for two different characteristic decay lengths, \(\lambda = 0.5\)~m (left panel) and \(\lambda = 1.3\)~m (right panel). Using red dashed lines, we plot the `fake' efficiency distribution that assumes all the outgoing particles decay at a distance equal to the characteristic decay length of the particle, i.e., assuming a Dirac-delta decay distribution in place of the exponentially falling decay distribution. We also indicate, using dashed lines, the geometric cuts on the angular distribution, as imposed in our analysis. 
    }
    \label{fig:eff_dist}
\end{figure}

In an attempt to interpret the VRE distribution in \cref{eq:fin_eff}, we plot it for two different values of \(\lambda\) in \cref{fig:eff_dist}. In both the panels, the solid black line indicates the VRE distribution as obtained from \cref{eq:fin_eff}. The dashed red line indicates the efficiency distribution if we had assumed all decays to happen at \(\lambda\) only, i.e., if we had used a Dirac-delta distribution, centred at \(\lambda\) in place of the exponentially falling distribution. As is intuitively clear from \cref{eq:fin_eff}, convoluting the exponential distribution in essence `smears' the distribution that we would have obtained by using a Dirac-delta. Take, for example, an \(A^\prime\) coming out at \(\vartheta = \pi/3\) with \(\lambda = 1.3\) m, as shown in the right panel of the figure. If the particle decayed at its characteristic length, it would have missed the geometric cuts described above. Hence, the probability of its detection would have been zero. Yet, as its decay is not deterministic, but given from a probability distribution, there is a possibility that it will decay within the fiducial volume, as reflected by the black curve. It is also to be noted that as the decay distribution is exponentially falling, it always peaks at \(l=0\). Therefore, particles with a characteristic decay length that is longer than covered in the fiducial volume, have a substantial probability of actually decaying within the said volume.

The VRE distribution gives the probability of detection of a displaced vertex in the direction \(\vartheta\) for a particular point in the \(M_{A^\prime}\) -- \(g^\prime\) space. The probability that an \(A^\prime\) is produced in the direction \(\vartheta\) for a particular point in the \(M_{A^\prime}\) -- \(g^\prime\) theory space, is given by the differential distribution \(d\sigma/d\!\cos\vartheta\). To get the net number of detected events for any given point in the (\(M_{A^\prime},\, g^\prime\)) theory space, we need to multiply the \(\vartheta\) dependent efficiency with the differential-cross section and integrate. In our computations, we prefer to work with the pseudorapidity,
\begin{align}
    \label{eq:eta_def}
    \eta&= -\ln\tan\left(\frac{\vartheta}{2}\right),
\end{align}
over \(\vartheta\). As the relationship between \(\vartheta\) and \(\eta\) is one-to-one, it is straightforward to translate the expression for efficiency in \cref{eq:fin_eff} to \(\eta\). With that done, the net number of accepted events corresponding to a particular channel, \(F\), of decay of the \(A^\prime\) is given by:
\begin{align}
    \label{eq:num_events}
    N_\mathrm{Tot}^F&= L_\mathrm{I} \times \int_{\eta_\mathrm{min}}^{\eta_\mathrm{max}}
    d\eta\;
    \frac{d\sigma}{d\eta}\times \mathrm{BR}_{F}\;\epsilon_F\; \mathcal{E}(\eta)\; 
    \Theta\!\left(p_T^\mathrm{min}-\sin\vartheta(\eta)\,|\vec{p}|\right),
\end{align}
where $N_\mathrm{Tot}^F$, \(L_\mathrm{I}\), \(\mathrm{BR}_F\), and \(d\sigma/d_\eta\) are the total number of events, the integrated luminosity, the branching ration to \(F\), and the differential cross section with respect to the pseudo-rapidity, respectively.  The end points of the \(\eta\) integral are determined by the geometric cuts given in \cref{eq:fid_vol_eta}, which roughly correspond to \(-1.3<\eta<1.7\). The factor \(\epsilon_F\) is the overall efficiency of identification of a final state \(F\) at the detector. We exclusively discuss lepton final states in this work, and for the electron and the muon, the efficiencies are:
\begin{align}
    \label{eq:pid}
    \epsilon_e&=0.93;\quad\quad \epsilon_\mu=0.86.
\end{align}
In \cref{eq:num_events}, we have also made the \(p_T\) cut, as given in \cref{eq:pt_cut}, explicit. Note that \(|\vec{p}|\) is determined by the mass of the particle, from energy conservation:
\begin{align}
    \label{eq:modp}
    |\vec{p}|&= \frac{s-M_{A^\prime}^2}{2\sqrt{s}}.
\end{align}

In the next section, we use \cref{eq:num_events}, which contains not only the production cross section and the branching ratio, but also the efficiency distributions and geometric and kinematic cuts, to bound the parameter space of an \(A^\prime\). We compute the actual cross section of production and the decay widths and branching fractions of the \(A^\prime\) to do so.

\section{%
    Case study: The dark photon 
    \label{sec:cs}
}

\begin{figure}[htpb]
    \centering
    \includegraphics[scale=1]{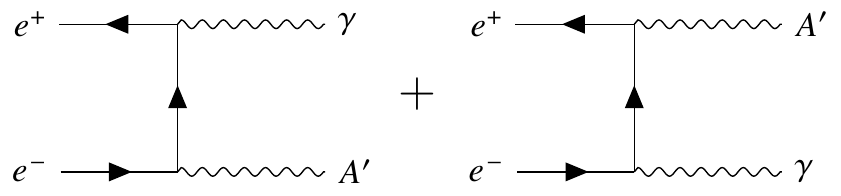}
    \caption{Diagrams contributing to the process under consideration.}
    \label{fig:prod}
\end{figure}

We now use the techniques discussed above in order to constrain the mass--coupling parameter space of the dark photon. We are interested in \(e^+e^-\to \gamma A^\prime\), i.e., the \(A^\prime\) is produced along with a photon in the \(t\)-channel annihilation of electron positron pairs. The relevant interaction Lagrangian is:
\begin{align}
    \label{eq:prod_lag}
    \mathcal{L}&= g^\prime A^\prime_\mu \bar{f}\gamma^\mu
    (g_v^e-g_a^e\gamma_5) f\,,
\end{align}
where \(A^\prime\) is a generic neutral vector boson that has both vectorial, \(g_v^e\), and axial, \(g_a^e\), couplings to the electrons (\(f)\). The \(A^\prime\) couples vectorially for all the cases we study, therefore, we drop the \(g_a^f\)  in all computations, to avoid confusion. The relevant diagrams for this process are given in
\cref{fig:prod}.

The corresponding spin-summed (averaged) amplitude-squared is:
\begin{align}
    \label{eq:amp_sq}
   \overline{|\mathcal{A}|^2}&= 8 \pi\, \alpha_\mathrm{EM}\; {g^\prime}^2
   {(g_v^e)}^2\, 
        \frac{(s+t)^2+(s+u)^2}{u t}\,.
\end{align}
Here, \(s, t, u\), and \(\alpha_\mathrm{EM}\) have their usual meanings. Given this amplitude-squared, the differential cross section for the process, in the CoM frame and in the limit of negligible electron mass, is:
\begin{align}
    \label{eq:dsec_eta}
    \frac{d\sigma}{d\eta}&= 
          \frac{\alpha_\mathrm{EM}}{2s}
          {g^\prime}^2 ({g_v^e})^2\;
          \frac{\cos^2\vartheta(\eta)(s-M_{A^\prime}^2)^2+(s+M_{A^\prime}^2)^2}
          {s(s-M_{A^\prime}^2)}.
\end{align}
It is this expression that we plug into \cref{eq:num_events} to compute the number of events for a particular \(M_{A^\prime},\,g^\prime\) pair. The form of the differential cross section is the same for all models we discuss, only the parameter \(g_v^e\) is model specific. 

We also need the information about the total and partial decay widths of the \(A^\prime\) in order to calculate the characteristic decay length and also to compute the branching ratios to different final states. The partial width to different channels for \(M_{A^\prime}\gtrsim \Lambda_\mathrm{QCD}\) is straightforward to compute using the effective Lagrangian given in \cref{eq:prod_lag}: 
\begin{align}
    \label{eq:dw}
    \Gamma_{A^\prime\to f\bar{f}}&=
        \frac{  n_c^f}{12\pi} {g^\prime}^2{(g_v^f)}^2 M_{A^\prime}
        \sqrt{1-4\frac{m_f^2}{M_{A^\prime}^2}}\;
        \left(1+2\frac{m_f^2}{M_{A^\prime}^2} \right),
\end{align}
with \(n_c= 1, 3\)  for leptons and quarks respectively. However, for \(M_{A^\prime}\sim \Lambda_\mathrm{QCD}\) and below, we need to compute the decay width to final state hadrons, for which we need to know the effective Lagrangian in question. To compute the decay width in this mass region, a variety of approximate techniques are used to get numbers of respectable accuracy (see, e.g., \cite{Tulin:2014tya,Bauer:2018onh}). In this work, we use the results of the data-driven approach, as given in Ref. \cite{Ilten:2018crw}, to get the widths. When discussing specific models below, we plot the variation of the partial widths and the branching  ratios as a function of \(M_{A^\prime}\).

\subsection{Limits for the dark photon}
\label{sec:prod_dec}

\begin{figure}[t]
    \centering
    \includegraphics[scale=1]{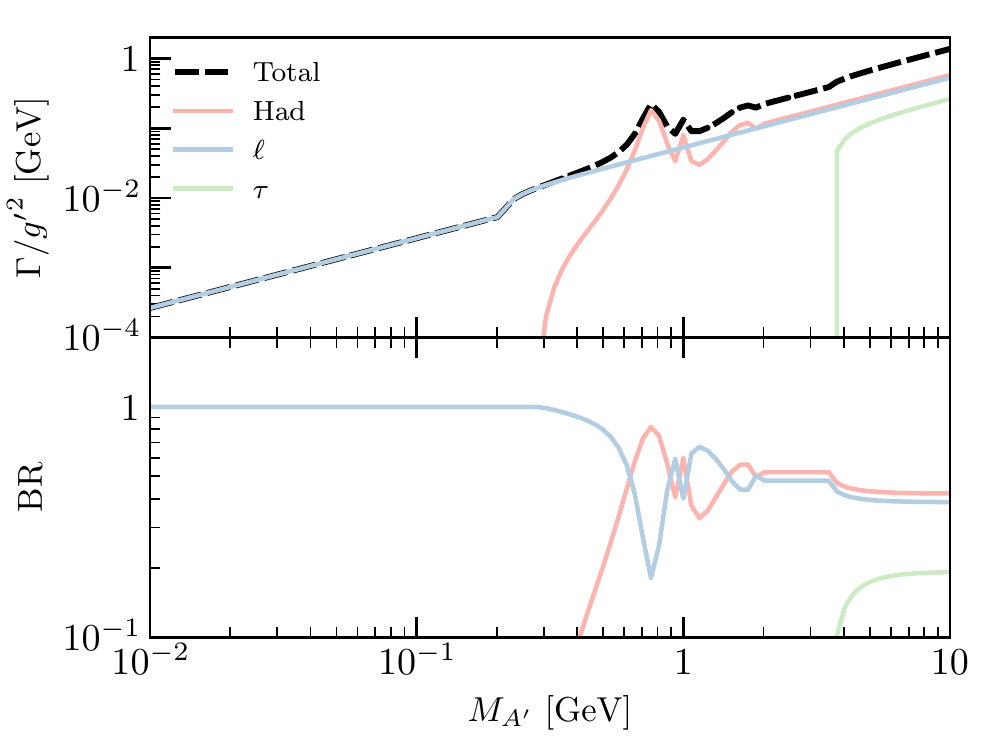}
    \caption{
        We plot the partial widths and the branching ratios of the dark photon for different channels of decay, as denoted in the legend.  The black dashed line in the top panel indicates the total width of the particle. We have plotted a scaled version of the decay width to remove the dependence on the coupling. 
    }
    \label{fig:DP_WBR}
\end{figure}

With the estimate for detector efficiencies and the differential cross section in hand, we can now derive the projected limits for the dark photon. The dark photon coupling to the SM fermions are generated through its gauge kinetic mixing with the photon \cite{Holdom:1985ag}. In the `original' basis (hatted), where the kinetic sector of the photon and the \(A^\prime\) is not canonically diagonalized, the Lagrangian for the dark photon is given by:
\begin{align}
    \label{eq:DPlag}
    \mathcal{L}&= -\frac{1}{4} \hat{F}^\prime_{\mu\nu} {\hat{F}^\prime}_{\mu\nu} 
        -\frac{\epsilon}{2} \hat{F}^\prime_{\mu\nu} \hat{F}^{\mu\nu}\,.
\end{align}
Here, \(\hat{F}_{\mu\nu}= \partial_\mu \hat{A}_\nu - \partial_\nu\hat{A}_\mu\) and \(\hat{F}^\prime_{\mu\nu}= \partial_\mu \hat{A}^\prime_\nu - \partial_\nu \hat{A}^\prime_\mu\) are the field-strength tensors corresponding to the photon and the dark photon fields respectively. To go to the canonically orthonormalized basis, we perform the following non-unitary transformation on the photon and the dark photon fields:
\begin{align}
    \label{eq:DPKR}
    \begin{pmatrix}
        \hat{A}_\mu \\ \hat{A}_\mu^\prime
    \end{pmatrix}
    &= \begin{pmatrix} 
        1 & -\epsilon\\
    0 & 1 \end{pmatrix}
        \begin{pmatrix}
        A_\mu \\ A_\mu^\prime
    \end{pmatrix}.
\end{align}
As a result of this transformation, the dark photon in the diagonal basis couples to all the photon currents with a strength that is proportional to the electric charges:
\begin{align}
    \label{eq:DP_coup}
    \mathcal{L}_\mathrm{DP}&=  e\epsilon \sum_f q_f A^\prime_\mu \bar{f}\gamma^\mu
    f\,  ,
\end{align}
where \(q_f\) is the electric charge of the fermion \(f\) and \(e\) is the QED coupling. Note, for the DP case, the effective \(A^\prime\) coupling strength is \(g^\prime= \epsilon e\). It is straightforward to check that this basis transformation keeps the photon mass, in the new basis, equal to zero. From \cref{eq:DP_coup} we can clearly see that \(g_v^e=-1\) for the dark photon, determining its differential cross section and its partial widths to different channels. In \cref{fig:DP_WBR}, we plot the partial widths of the dark photon along with the branching fractions for the available channels, as functions of the dark photon mass, \(M_A^\prime\). As discussed above, we use the results given in Ref. \cite{Ilten:2018crw} for the partial width to hadrons for \(M_A^\prime\lesssim \Lambda_\mathrm{QCD}\). We have cross-checked the result with the partial width given in a SHIP note\footnote{\url{https://cds.cern.ch/record/2214092/files/ship-note-dark-photons.pdf}}.

\begin{figure}[t]
    \centering
    \includegraphics[scale=0.7]{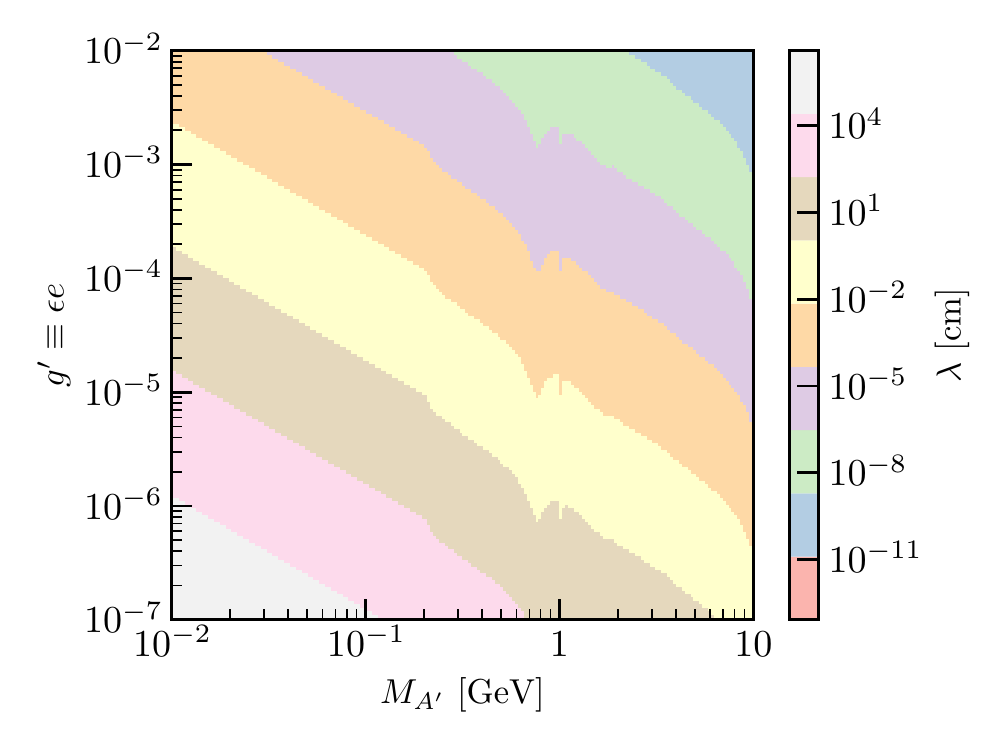}
    \includegraphics[scale=0.7]{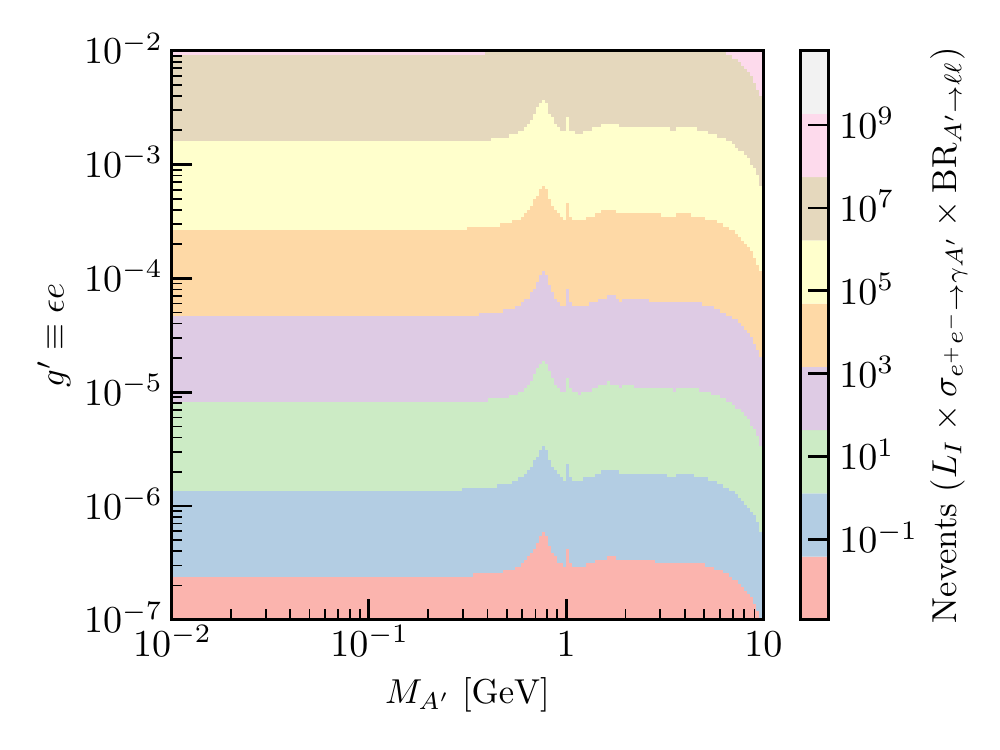}
    \caption{
        Left: The characteristic length of decay, \(\lambda\), for different values of \(M_{A^\prime}\) and \(g^\prime (\equiv \epsilon e)\). The shades indicate different ranges for \(\lambda\), as given by the colorbar to the right. Right: We plot the number of events produced at 50~ab\textsuperscript{-1} of luminosity for different values of \(M_{A^\prime}\) and \(g^\prime\). The shades indicate different ranges for the number of events produced, as given by the colorbar to the right. 
    } 
    \label{fig:nevents_dp}
\end{figure}

Given the width and the differential cross section of the \(A^\prime\), we obtain its decay length distribution from the characteristic length, \(\lambda=\beta\gamma c\tau\). The boost factor \(\beta\gamma\) can be obtained from the magnitude of the three momentum given in \cref{eq:modp} as \(\beta\gamma = |\vec{p}|/M_{A^\prime}\). In the left panel of \cref{fig:nevents_dp}, we plot contours indicating the characteristic length in the (\(M_{A^\prime},\, g^\prime\)) plane. In the right panel of the figure, we give contours for the total number of events, \(N_\mathrm{events}\), in the same (\(M_{A^\prime}, g^\prime\)), to estimate the number of events produced. The number of events are calculated for an integrated luminosity of 50~ab\textsuperscript{-1} and for \(\ell\equiv e+\mu\). The cuts on the \(p_T\) and \(\eta\), as discussed in the previous section, are applied in calculating the number of events. However, no requirement on the displacement of the \(A^\prime\) decay vertex is imposed for this plot, i.e., this is the number of events before the convolution by the VRE distribution. We do this to show the variation of the cross section with the mass and the coupling. Below, where we draw the actual exclusion limit, we use the cross section convoluted by the VRE distribution, as given in \cref{eq:num_events}.

\begin{figure}[t]
    \centering
    \includegraphics[scale=1]{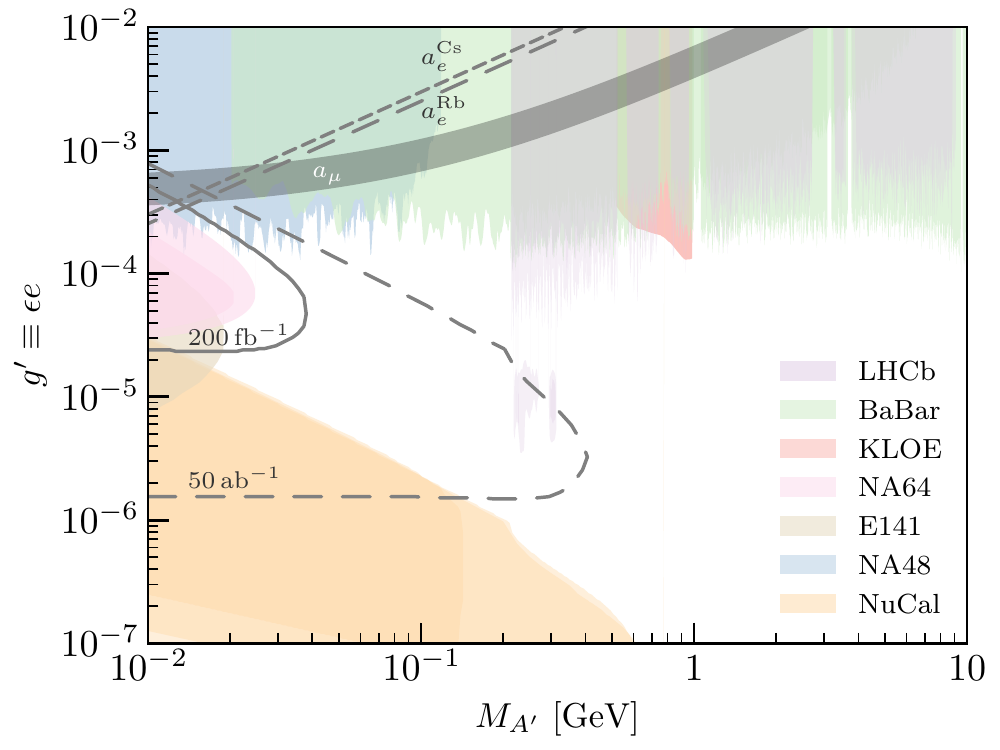}
    \caption{The dashed grey line demarcates the outer boundaries of the projected limits for Belle II sensitivity to displaced searches for the dark photon model at an integrated luminosity of 50~ab\textsuperscript{-1}, with the dark photon decaying to a lepton pair (\(e\) or \(\mu\)). The solid grey line indicates the same but for an integrated luminosity of 200~fb\textsuperscript{-1}. The regions shaded in different colors reflect bounds from existing experiments, with the colors used for different experiments indicated in the legend. The regions of parameter space consistent with the latest measurement of \(\Delta a_\mu\) has been shaded in grey and labelled. Also shown are the contours which satisfy the \(a_e\) measurements using Rb and Cs.}
    \label{fig:moneyplot_DP}
\end{figure}

With all the relevant information in hand, we can finally present the actual limits on the parameter space from displaced vertex searches. To get the limits, we accept those (\(M_{A^\prime},\,g^\prime\)) points for which the number of accepted events, \(N\), passes our acceptance criteria. If we assume Poisson statistics, the number of accepted events need to be greater than 2.3(\(\sim 3\)) for a 90\% C.L. exclusion limit, assuming no background. In \cref{fig:moneyplot_DP}, we draw the exclusion limits for integrated luminosities of 50~ab\textsuperscript{-1} (the design luminosity of Belle II) and 200~fb\textsuperscript{-1} (approximately the data collected by the collaboration as of writing). We draw projections for the final state of leptons (\(\ell\equiv e+\mu\)).  From \cref{fig:moneyplot_DP}, we can see that once Belle II collects its machine specification of 50~ab\textsuperscript{-1} of data, it would exclude a large part of the \(M_{A^\prime},\, g^\prime\) parameter space for \(10\) MeV \(< M_{A^\prime} < 500\) MeV, the region of sensitivity being demarcated by the dashed contour. It is noteworthy that this region of parameter space, that the Belle II displaced vertex searches are sensitive to, nicely complements existing limits. We have set a lower limit of \(M_{A^\prime}\)=10~MeV to stay  clear of the limits set on an electrophilic light particle from BBN and CMB observables \cite{Sabti:2019mhn}. For the limit at 50~ab\textsuperscript{-1}, a coupling below \(g^\prime\sim 10^{-6}\), gives a production cross section too small to produce enough displaced vertex signatures. In the other direction, both large masses and coupling strengths make the characteristic decay length too small for the vertex to fall in our region of acceptance. These observations can also be intuited from the two panels in \cref{fig:nevents_dp}. We note that even with the data collected as of now by Belle II, it is expected to have surpassed the limits set by existing experiments in the 10~MeV \(< M_{A^\prime} < \) 50~MeV, \( 2\times 10^{-5} < g^\prime < 5\times 10^{-4}\). The existing bounds in this region come from the NA64 collaboration and the E141 collaboration, as discussed below. The projected sensitivities that we show are determined by \(r_\mathrm{min/max}\) and \(z_\mathrm{min/max}\), as discussed in \cref{sec:eff}. We have checked that the sensitivities are robust against moderate changes in the limits of \(z\). We do not consider changes in the limits of \(r_\mathrm{min/max}\), as the track reconstruction efficiency that we consider depends directly on \(r_\mathrm{min/max}\) through the normalizing factor, as discussed in \cref{sec:eff}.

Along with the projected limits from Belle II, in \cref{fig:moneyplot_DP} we also plot existing limits on the DP parameter space from different experiments. A list of these experiments, along with short descriptions, are given in \cref{sec:intro}. We have used \texttt{DarkCast}\footnote{\url{https://gitlab.com/philten/darkcast}, GNU GPL V2.}, the companion package of Ref. \cite{Ilten:2018crw}, to recast the limits from existing experiments to the \(A^\prime\) parameter space. For the existing bounds, we have only shown those limits which are the strongest. The bounds displayed are listed below. In red, we show the bounds reported by the KLOE experiment \cite{KLOE-2:2018kqf}. Bounds coming from prompt and displaced (the `islands' in the middle of the plot) searches by the LHCb collaboration~\cite{LHCb:2017trq, LHCb:2019vmc} have been shown in violet. The bounds from the BaBar experiment (prompt)~\cite{BaBar:2014zli, BaBar:2016sci, BaBar:2017tiz}, covering most of the mass region in consideration, have been given in green. The NA48/2 collaboration's search for dark photons in \(\pi^0\) decays gives the  strongest bound below the \(\pi^0\) mass \cite{NA482:2015wmo}, as shown in blue. The strongest constraint from existing displaced searches \cite{Tsai:2019buq} comes from the \(\nu\)-Calorimeter I (NuCal) experiment \cite{Blumlein:1990ay, Blumlein:1991xh}, as shown in orange. We have not shown the bounds coming from other displaced vertex searches, e.g., by the E137 collaboration \cite{Bjorken:2009mm,Andreas:2012mt} or that at Orsay \cite{Davier:1989wz}, as the NuCal bound supersedes the bounds coming from both these experiments. In brown, we have given the bounds from the displaced searches by the E141 collaboration \cite{Riordan:1987aw}. The pink region  denotes the bound obtained by the NA64 experiment~\cite{NA64:2019auh} while searching for the hypothetical X(17) particle.

In \cref{fig:moneyplot_DP}, we have also shaded in grey the band that denotes the 2$\sigma$ limit that is consistent with the anomalous muon magnetic moment, as obtained in Ref. \cite{Bodas:2021fsy}. Using dashed lines, we also denote the contours in the parameter space which are consistent with the \((g-2)_e\) measurements using the fine structure constant measurements from the Caesium \cite{Parker:2018vye,Aoyama:2017uqe} and Rubidium \cite{Davoudiasl:2018fbb, Morel:2020dww} atoms. We observe that the regions of parameter space that are consistent with these measurements are already ruled out by the BaBar and the NA48/2 collaborations and Belle II is sensitive to the coupling strengths corresponding to these measurements only for a very small mass range~(\(\lesssim\)~20~MeV). 

Before concluding this section, two comments are in order. First of all, in general, the DP also mixes kinetically with the \(Z\) boson that induces couplings proportional to the \(Z\) couplings to the fermions. In this work, we do not consider \(A^\prime\)--\(Z\) mixing of any form, only to simplify the presentation of our results. Note, we do not invoke the hierarchy of the \(A^\prime\) and \(Z\) mass scales when ignoring this mixing. Despite a hierarchical separation, other Lagrangian parameters can be tuned to have sizeable mixing (see, e.g., Ref. \cite{Bandyopadhyay:2018cwu}). Also, we are giving bounds on couplings which are extremely small \(\sim 10^{-6}\). So, even when the mixing is of the order of \(M_{A^\prime}^2/M_Z^2\), it would still be commensurate with the couplings we are giving bounds on.  Secondly,  we generally invoke dark photons as portals between the SM and some dark sector. Therefore, the typical dark photon has couplings to dark sector particles.  The presence of a dark photon decay channel to dark sector particles in effect only adds an additional contribution to the total width of the dark photon, scaling the branching ratios and the characteristic decay length of the dark photon. The modification to the width and branching ratios depend on the masses of the dark sector particles and makes the analysis model dependent. Hence, for the sake of benchmarking, we do not include any dark sector particles in our work. Modifications to decay widths due to dark sector particles and modifications to coupling strengths due to \(A^\prime\)--\(Z\) mixing etc.\ can easily be introduced in an analysis that recasts these results to a different model. 

\section{Models of gauged \texorpdfstring{\(Z^\prime\)}{Z'}}
\label{sec:model_zprime}

Models where the anomaly free gauge symmetries of the SM are gauged, viz. \(L_\mu-L_\tau\), \(L_e-L_\mu\), \(L_\tau-L_e\), and \(B-L\)\footnote{It should be noted, anomaly cancellation in the \(B-L\) model requires the addition of an RH neutrino per generation if the gauge-gravity mixed anomaly is considered.}, are extremely popular due to a variety of reasons, some of which we have mentioned in the introduction. In this section, we discuss the limits on these models and the model of the `protophobic' force (see Ref. \cite{Feng:2016ysn}, also discused below). Among these, the \(L_\mu-L_\tau\) model stands out by virtue of not interacting with the first generation of fermions at leading order. This also means that its production at electron colliders is suppressed, giving signatures different from the rest. Hence, we discuss the \(L_\mu-L_\tau\) model in some detail, before moving on to the other models.

\subsection{\texorpdfstring{\(Z^\prime\)}{Z`} 
    with \texorpdfstring{\(L_\mu-L_\tau\)}{Lmu-Ltau} symmetry
    \label{sec:models}
}
In \(L_\mu-L_\tau\) models, interactions to the first generation fermions are generated through kinetic mixing between the \(Z^\prime\) boson corresponding to the gauged \(U(1)_{L_\mu-L_\tau}\) and the SM photon. This allows us to give projections on variants of this model from the Belle II experiment. 

\begin{figure}[t!]
    \centering
    \includegraphics[scale=0.7]{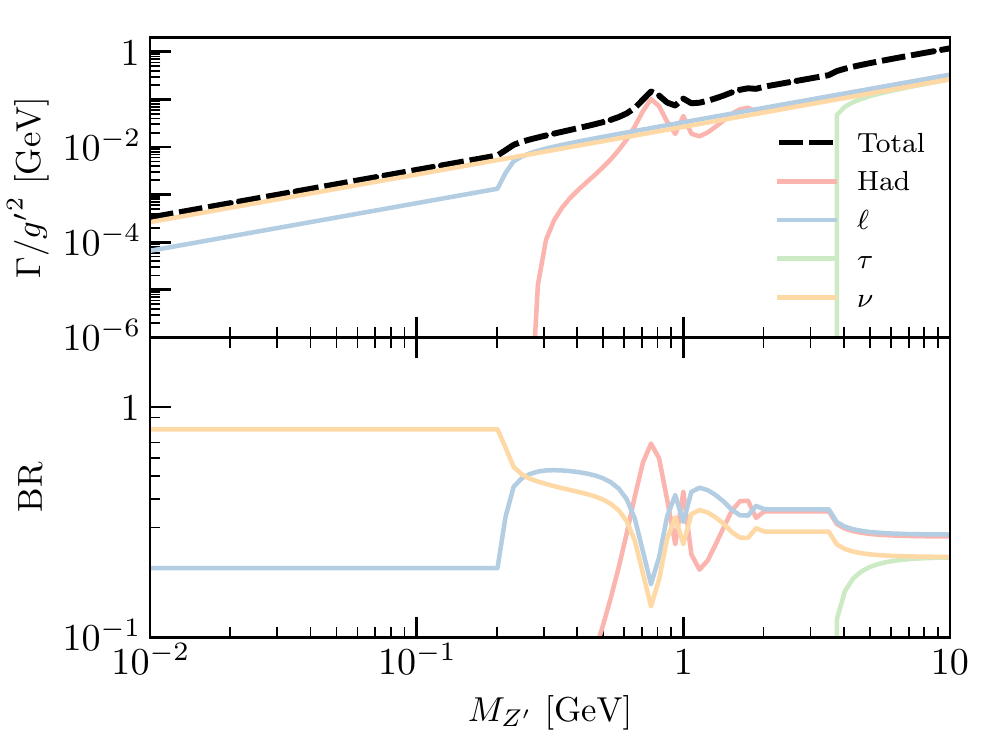}
    \includegraphics[scale=0.7]{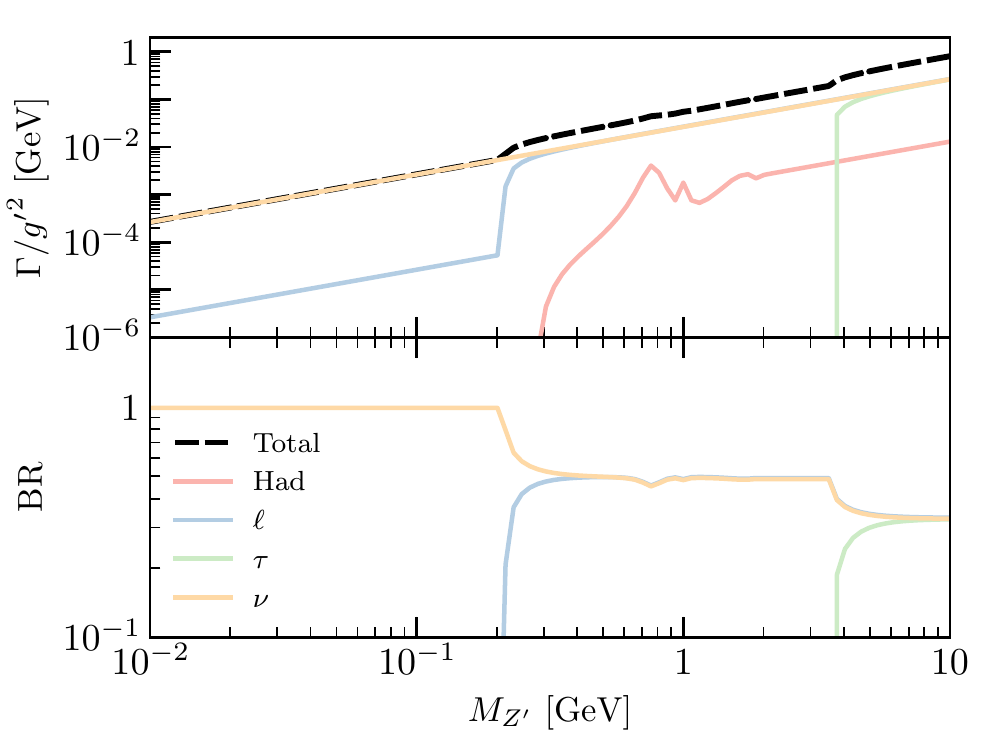}
    \caption{
    Partial decay widths and branching ratios for the \(L_\mu-L_\tau\) \(Z^\prime\) boson. Left:  for \(r_f = 0.5\). Right: for \(g_v^e/g_v^\mu= 0.1\). See text for details.}
    \label{fig:BR_mt}
\end{figure}

\begin{figure}[t!]
    \centering
    \includegraphics[scale=0.7]{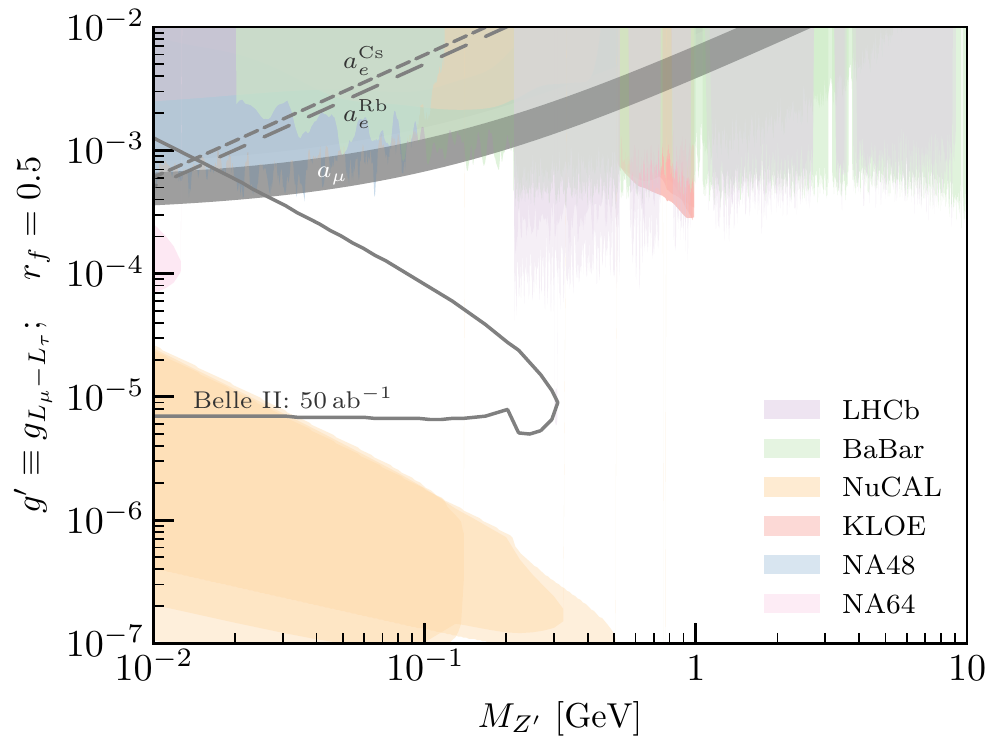}
    \includegraphics[scale=0.7]{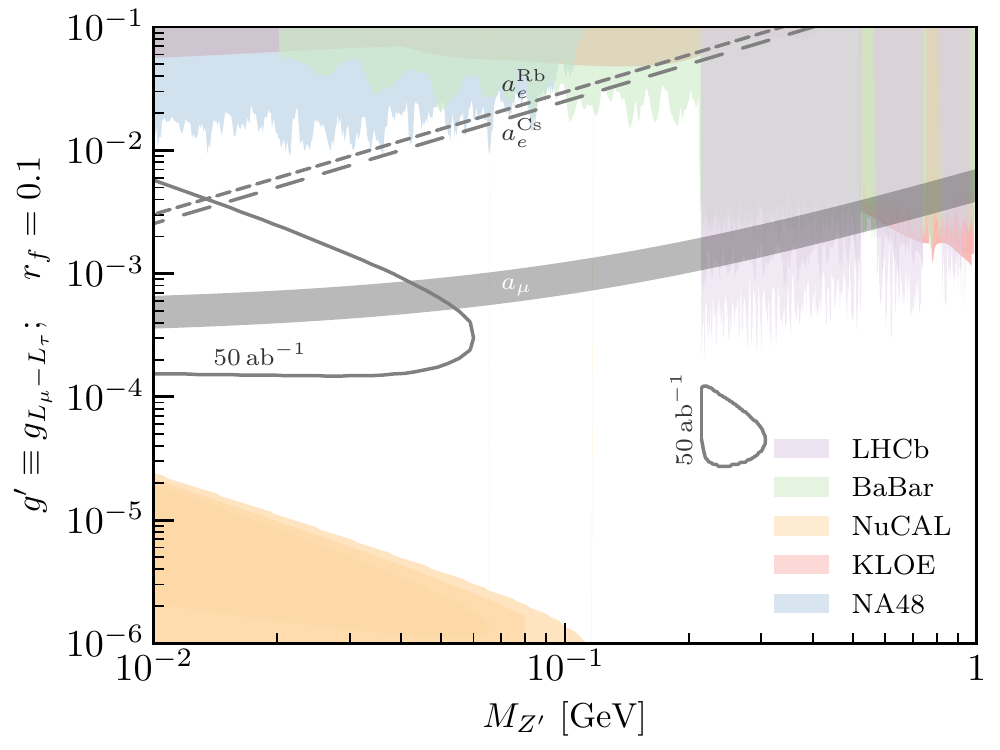}
    \caption{
    Similar to \cref{fig:moneyplot_DP}, the black contour can be probed with displaced vertex searches at Belle II for a $L_{\mu}-L_\tau$ gauge boson with $r_f=0.5$ (left) and $r_f=0.1$ (right). As is evident for the \(r_f=0.1\) case, smaller values of the electron coupling reduces the production and hence relaxes the bound. 
}
    \label{fig:moneyplot_mutau}
\end{figure}

In the basis where the gauge kinetic sector is not diagonal, the \(Z^\prime\) couples to the fermions with a charge proportional to the difference between the muon lepton number and the tau lepton number. The corresponding interaction Lagrangian is given by:
\begin{equation}
    \mathcal L_{Z^\prime} = g_{L_\mu-L_\tau} 
        \left(\bar\ell_\mu\gamma^\alpha \ell_\mu+\bar\mu_R\gamma^\alpha\mu_R - 
            \{\mu \rightarrow\tau\} 
        \right) Z^\prime_\alpha\;,
\end{equation}
where $\ell_\mu$ is the left-handed muon doublet.  However, there is a gauge kinetic mixing term, between the \(Z^\prime\) and the photon, of the form:
\begin{equation}
    \mathcal L_{Z^\prime} \supset \frac{\epsilon}{2} Z^\prime_{\mu\nu}F^{\mu\nu}\;.
\end{equation}
Being a marginal operator and consistent with gauge symmetries, the kinetic mixing term would be present unless special discrete symmetries are introduced such as $\mu\leftrightarrow\tau$ and $Z^\prime\to-Z^\prime$~\cite{Dienes:1996zr,Banerjee:2018mnw}. However, even then, with fermions charged under both electromagnetism and \(L_\mu-L_\tau\), kinetic mixing will be generated at one loop. The generic form of the resulting mixing is
\cite{Cheung:2009qd}:
\begin{align}
    \label{eq:km_gg}
    \epsilon&= \frac{e g_{L_\mu-L_\tau}}{16\pi^2}\sum_f q_f x_f \ln\left(\frac{\mu^2}{M_f^2}\right)\;,
\end{align}
where \(x_f\) and \(M_f\) are the charge and the mass of the fermion \(f\) in the loop, and \(\mu\) is the scale of renormalization. Clearly, with many fermions (not necessarily from the SM spectrum) in the loop and with large logs, this number can be quite large, around \(10^{-1}\) (see, e.g., Ref. \cite{Gherghetta:2019coi}). In this work, we take a model agnostic approach and treat the kinetic mixing as a free parameter. This term induces universal couplings between $Z^\prime$ and the SM fermions, including electrons, in a way exactly similar to the dark photon case. We take a hierarchical separation between the coupling of the \(L_\mu - L_\tau\) boson to muons and taus, as compared to all
other fermions:
\begin{align}
    |g_v^f|&= r_f\, |g_v^{\mu/\tau}|,\; \forall f\notin \{\mu, \tau\}\,,
\end{align}
where \(g_{L_\mu-L_\tau}\cdot\, g_v^f\) is the coupling of the \(Z^\prime\) boson to the \(f\) fermion. In \cref{fig:BR_mt}, we plot the branching ratios and partial widths for \(r_f=0.5\) and \(r_f=0.1\). In \cref{fig:moneyplot_mutau}, we show the limits for the same two values of \(r_f\). As is evident from the figure, the former case is similar to the dark photon. However, the latter shows interesting deviations from the dark photon limits and merits additional comments.

For the \(r_f = 0.1\) case, the experiment is sensitive to two disjointed regions of the parameter space. We discuss the contour in the lower mass regions first, and then move on to the `island'. The sensitivity region for \(M_{Z^\prime}\lesssim 80\) MeV shows deviations from the dark photon case for two reasons. First of all, the production from \(e^+ e^-\) is governed by the coupling strength to electrons, therefore, is suppressed by \(r_f^2\).  As a result, the bound starts from larger coupling values compared to the dark photon case. Also, in this region, decay to muons is kinematically forbidden and that to electrons is controlled by the same sub-dominant \(r_f^2\). Hence, the dominant branching is to neutrinos. This can also be seen from the right panel of \cref{fig:BR_mt}, where we plot the partial widths and branching ratios for this model. 
Notice that the sensitivity for the \(r_f=0.1\) case extends to larger values of \(g^\prime\), compared to the DP case (or the \(r_f=0.5\) case). The reason is two fold. Firstly, the coupling to electrons is suppressed by the additional \(r_f\) factor. This implies, that we need larger \(g^\prime\) values to get a cross section equal to the DP case. Secondly, the \(r_f\) suppression  to the coupling leads to suppressed decay lengths. Hence, for larger values of \(g^\prime\) the decay length is still large enough for displaced searches to be sensitive.  The most interesting consequence of this is that the Belle II sensitivity region ends up covering a large region of the parameter space that is compatible with the \((g-2)_\mu\) anomaly. We note, unlike the DP case, where the \(a_\mu\) region is mostly ruled out by NA48 \cite{NA482:2015wmo} and BaBar exclusions \cite{BaBar:2014zli,BaBar:2016sci, BaBar:2017tiz}, it is the Belle II exclusion that has the potential to rule out most of the \(a_\mu\) region for this model. Finally, we see that the experiment is sensitive to a small island of parameter space just above the 2\(\mu\) threshold. The reason for this is that at this threshold the cross section times branching ratio to two leptons gets an enhancement due to the opening up of the muon channel for the \(Z^\prime\) to decay to. As a result, displaced vertex searches become sensitive for a small slice of mass before losing sensitivity as the inclusion of the \(\mu\) channel increases the total decay width to an extent that displaced vertex searches become untenable. 

\subsection{Other Models}
\label{sec:other_models}

In this subsection, we discuss the projections for displaced vertex searches at Belle II for a few other models of interest. As the analysis is the same as that discussed before, we just discuss the limits for these models.  The models we discuss are the \(L_e-L_{\mu/\tau}\) models, the \(B-L\) model, and the protophobic model. As all of these models couple to first generation fermions, the limits are similar across them, with minor differences that we discuss in brief. 

\begin{figure}[h]
    \centering
    \includegraphics[scale=0.7]{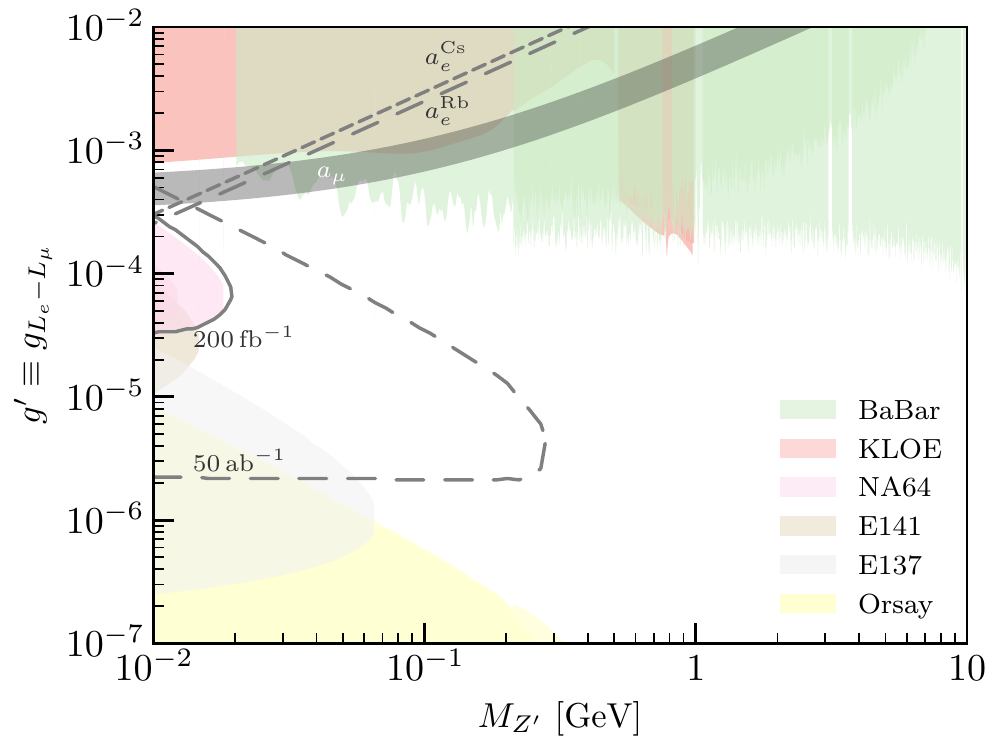}
    \includegraphics[scale=0.7]{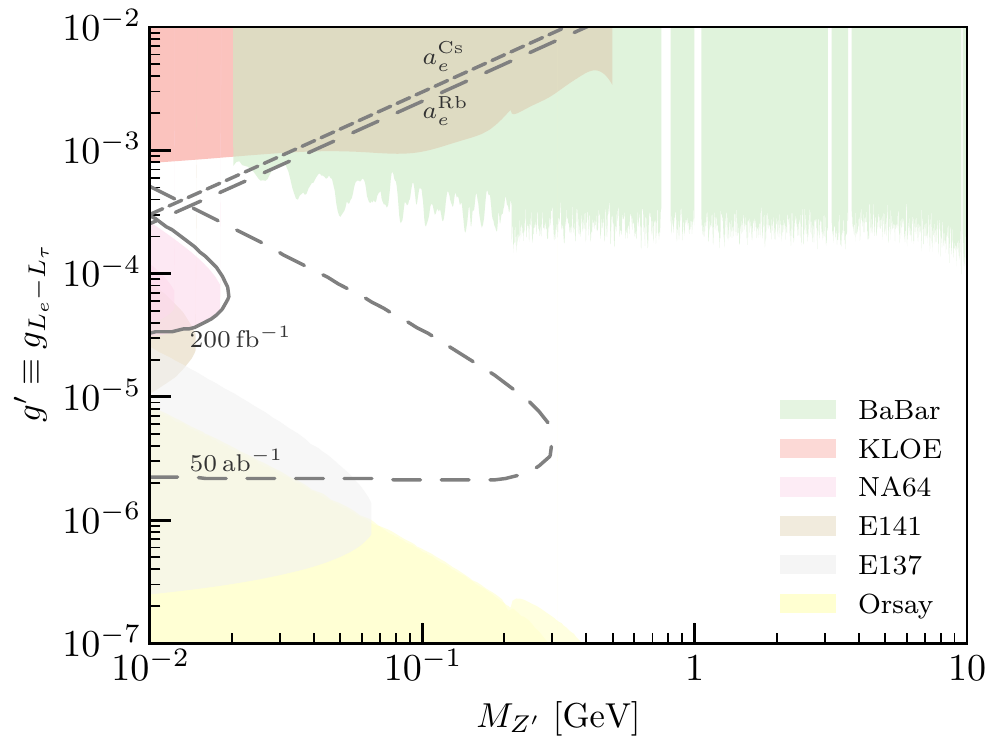}
    \includegraphics[scale=0.7]{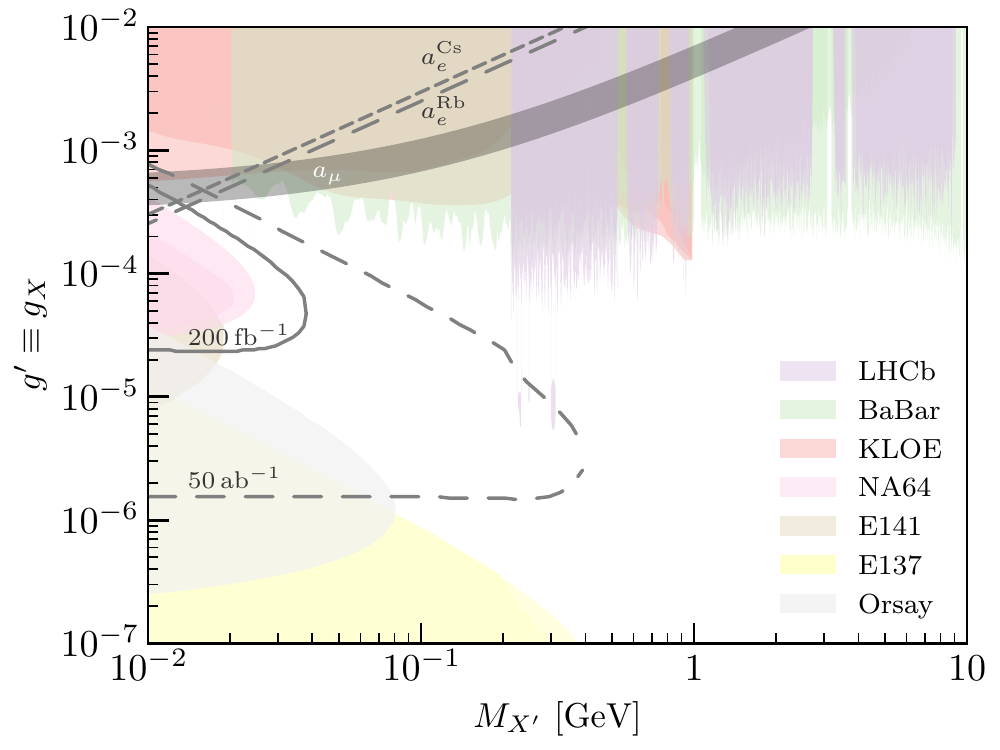}
    \includegraphics[scale=0.7]{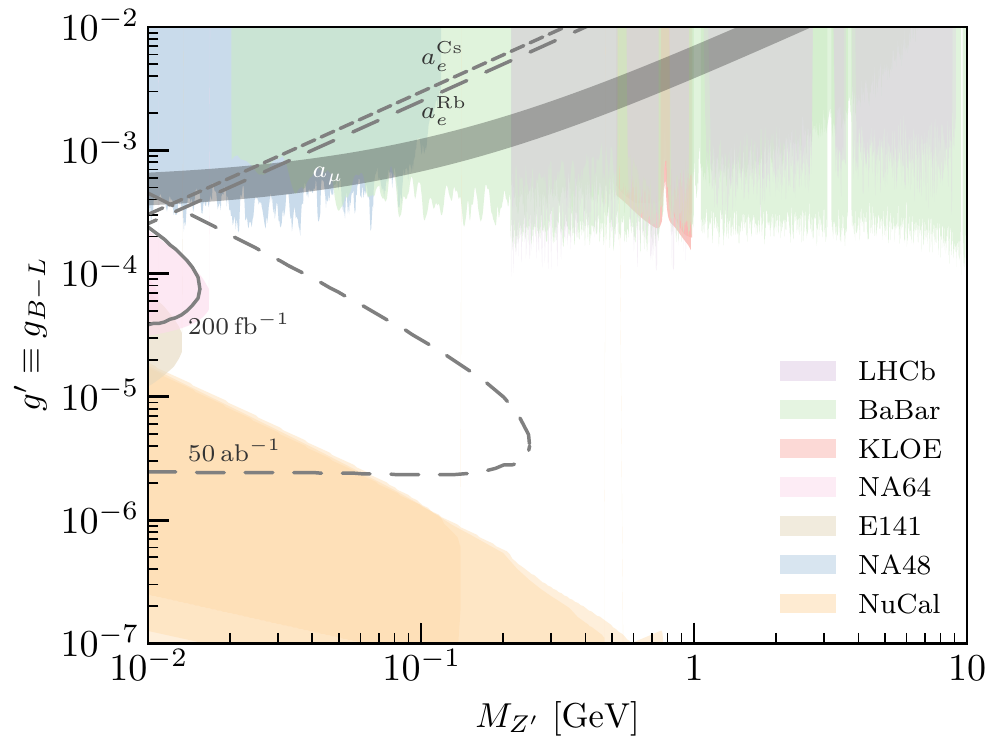}
    \caption{In the top left and right panels, we plot the projected limits in the parameter space of the \(L_e-L_\mu\) and the \(L_e-L_\tau\) models respectively. In the bottom left panel and the bottom right panels, we draw the same for the protophobic model and the \(B-L\) model respectively. 
    }
    \label{fig:moneyplot_BL}
\end{figure}

In the top two panels of \cref{fig:moneyplot_BL}, we plot the Belle II projections and existing limits for the \(L_e-L_\mu\) model (left) and the \(L_e-L_\tau\) model (right). In these models, the gauge boson couples to the electron at leading order. As a result, the projected limits extend for a large portion of the parameter space. Note, we do not assume sizeable kinetic mixing for these models. Needless to say, large kinetic mixing coefficients can be generated for these models too, in which case these projections will be modified. The existing bounds for the both the \(L_e-L_i\) models are somewhat different from the dark photon case. As the \(Z^\prime\) in these models do not interact with hadrons at leading order, there is no bound from the NuCal (proton beam) data. Instead, the strongest displaced vertex bounds, in regions of small coupling, come from the electron beam dump data as collected at Orsay \cite{Davier:1989wz} and by the E137 collaboration \cite{Bjorken:1988as}. Also, LHCb or NA48 data does not give limits on the parameter space of these models, with the large coupling region only constrained by BaBar \cite{BaBar:2014zli,BaBar:2016sci,BaBar:2017tiz} and KLOE \cite{KLOE-2:2018kqf}.

The protophobic and the  \(B-L\) models differ from the other two in the sense that in these models, the \(Z^\prime\) has \(\mathcal{O}(1)\) couplings to the quarks. The protophobic model was postulated as an explanation of the ATOMKI anomaly~\cite{Krasznahorkay:2018snd,Krasznahorkay:2021joi}. The gauge boson in the model, \(X(17)\), couples universally and vectorially to the three generations. The couplings to the up-type quarks, down-type quarks, charged leptons, and neutrinos are \(-1/3, 2/3, -1\), and 0 respectively. With these couplings, it is easy to see that the net charge of the proton is zero. In the bottom left panel of \cref{fig:moneyplot_BL}, we show the projected limits in the parameter space of this model. The important thing to note from the figure is that at 200~fb\textsuperscript{-1}, i.e., the amount of data Belle II is expected to have collected by now, the Belle II displaced limit gives a stronger bound than the limits set by the NA64 collaboration in a dedicated search of this particle~\cite{NA64:2019auh}, as given by the pink region. That is, an analysis of the present Belle II data against the displaced vertex hypothesis has the potential to reproduce the results of the NA64 collaboration and in the process rule out (or discover) the X(17) explanation of the ATOMKI anomaly. As for existing beam dump searches, although the X boson interacts with quarks, there is no strong bound from NuCal in this case, as the quark couplings of the X boson are engineered specifically to have no interaction with the proton. The strongest displaced constraints for the X particle comes from the electron beam experiments at Orsay and E137. Finally, in the bottom right panel of \cref{fig:moneyplot_BL}, we plot the projected limits for the \(B-L\) model for two integrated luminosities, \(L_I=200\) fb\(^{-1}\) and \(L_I=50\) ab\(^{-1}\). From the figure we see that the limit and the existing bounds are more or less similar to those for the dark photon. This is expected, as the couplings of the vector boson in the two models are similar.

\section{Conclusion}
\label{sec:conclusion}
Displaced vertex searches at the Belle II experiment has enormous potential to scrutinize a large part of the parameter space of models with a vector boson with mass \(\lesssim 0.5\) GeV. Our projections, both for the Belle II design luminosity of 50~ab\textsuperscript{-1} and the expected current luminosity of 200~fb\textsuperscript{-1}, show that the collaboration should consider analysing there data to test the displaced vertex hypothesis. Such searches not only have the possibility to probe as yet uncharted territories in the light vector boson parameter space, but also to test specific scenarios invoked as explanations to hints of new physics. Such scenarios include, but are not limited to, portals to dark matter, new vector bosons to solve the \((g-2)_\mu\) anomaly, and the protophobic boson postulated to solve the ATOMKI anomaly. 

\section*{Note Added}
While finishing our work, Ref.~\cite{Ferber:2022ewf} appeared which also looked at the displaced vertex searches for a dark photon at Belle~II. We present a more analytical and detailed estimation of the efficiency pertaining to displaced vertices. Our DP results overlap with the findings of Ref.~\cite{Ferber:2022ewf} for some regions of parameter space. We also extend the analysis for light $Z^\prime$ models where displaced vertex searches can probe significant part of the parameter space.

\acknowledgments
SC and ST was supported by the grant MIUR contract 2017L5W2PT. We thank Aleksandr Azatov for comments and careful reading of the manuscript. SC would also like to thank Heerak Banerjee and Sourov Roy for clarifications regarding Ref.~\cite{Banerjee:2018mnw}.

\bibliographystyle{JHEP}
\bibliography{dp}

\end{document}